\baselineskip=12pt

\magnification 1200
\vsize=8.5 truein
\hsize=5.8 truein
\hoffset=1.0 cm
\voffset=1.0 cm

\overfullrule = 0pt

\font\ftit=cmbx12

\parskip=6pt
\parindent=2pc

\font\titulo=cmbx10 scaled\magstep1

\def\qed{{\vrule height 7pt width 7pt depth 0pt}\par\bigskip}

\def\section#1{\vskip 1.5truepc plus 0.1truepc minus 0.1truepc
	\goodbreak \leftline{\titulo#1} \nobreak \vskip 0.1truepc
	\indent}
\def\frc#1#2{\leavevmode\kern.1em
	\raise.5ex\hbox{\the\scriptfont0 $ #1 $}\kern-.1em
	/\kern-.15em\lower.25ex\hbox{\the\scriptfont0 $ #2 $}}


\def\IR{{\rm I\!R}}  
\def\IN{{\rm I\!N}}  

\font\cmss=cmss10
\font\cmsss=cmss10 at 7pt

\def\IZ{\relax\ifmmode\mathchoice
{\hbox{\cmss Z\kern-.4em Z}}{\hbox{\cmss Z\kern-.4em Z}}
{\lower.9pt\hbox{\cmsss Z\kern-.4em Z}}
{\lower1.2pt\hbox{\cmsss Z\kern-.4em Z}}\else{\cmss Z\kern-.4em Z}\fi}

\centerline{\ftit DISCRETE CHAOS}

\vskip 1.5pc

\centerline{Henri Waelbroeck
{\footnote{*}{On sabbatical leave from the Institute of Nuclear Sciences, UNAM;
Circuito Exterior, C.U.; A. Postal 70-543; Mexico DF 04510}}}

\centerline {Instituut voor Theoretische Fysica}
\centerline{Universiteit Utrecht}
\centerline{Princetonplein 5, Postbus 80006}
\centerline{3508 TA Utrecht, The Netherlands}
\centerline{e-mail: waelbrk@phys.uu.nl}
 
\vskip 0.5pc
 
\centerline{Federico Zertuche}

\centerline{Instituto de Matem\'aticas, UNAM}
\centerline{Unidad Cuernavaca, A.P. 273-3, Admon.~3}
\centerline{62251 Cuernavaca, Morelos, M\'exico.}
\centerline{e-mail: zertuche@matcuer.unam.mx}

\vskip 1.5pc

\centerline {Abstract}  
{\leftskip=1.5pc\rightskip=1.5pc\noindent We propose a theory of
deterministic chaos for discrete systems, based on their
representations in binary state spaces $ \Omega $, homeomorphic to the
space of symbolic dynamics. This formalism is applied to neural
networks and cellular automata; it is found that such systems cannot
be viewed as chaotic when one uses the Hamming distance as the metric
for the space. On the other hand, neural networks with memory can in
principle provide examples of discrete chaos; numerical simulations
show that the orbits on the attractor present topological transitivity
and a dimensional phase space reduction. We compute this by extending
the methodology of Grassberger and Procaccia to $ \Omega $. As an
example, we consider an asymmetric neural network model with memory
which has an attractor of dimension $D_a = 2$ for $N = 49$.}

\

\noindent Short Title: {\it Discrete Chaos}


\vfill\eject

\vskip 1.5pc

\section{1. Introduction.}

	Recently, discrete systems with a complex dynamical behavior
have received a great deal of attention, for their relevance in
fields ranging from theoretical biology to quantum gravity. For
example, asymmetric neural networks~$ {}^{[1-4]}$ can have a
complicated dynamical behavior which is reminiscent of ``chaos''.  Also
cellular automata display~$ {}^{[5]}$ bifurcations between several
possible dynamical regimes~$ {}^{[6]} $, the most disordered of which
has been described as ``chaotic''~$ {}^{[7]} $.  Yet it is unclear
precisely how this type of dynamics in discrete spaces is related to
deterministic chaos in a Euclidean phase space. 

	In this article we will examine how the definitions of 
deterministic chaos can be translated to the context of discrete state
spaces. This will lead us to a formalism which we call ``discrete
chaos'', that allows one to decide whether or not the complex dynamics 
of some finite systems can be viewed as chaotic in the limit in which
the system grows to infinity. 

 Unfortunately, for most finite systems there is no convenient
quasi-represen\-tation in terms of real variables. For example, in
neural networks and cellular automata the relevant distance is the
Hamming distance; this induces a discrete topology on the space of
states that is distinct from the usual topology of $ \IR^n $. There
are different points of view on this problem, ranging from the
fundamentalist, which concludes that a finite system cannot be viewed
as approximately chaotic, to the liberal, which reduces the definition
of chaos to the exponential growth of the limit-cycle period with the
size of the system.

	Our feeling is that chaos should not be limited to real
variables, as these are idealizations of a reality which could be
viewed equally well in terms of finite state spaces. Indeed, the fact
that most real numbers have infinite algorithmic information~$
{}^{[8]} $ is not really satisfactory from a physicist's point of view. Yet
some form of idealization is necessary to define chaos rigorously.

	Our purpose in this article is to propose a different
idealization, inspired from symbolic dynamics~$ {}^{[9-13]} $. We will
assume that one is given a representation of the system through a
sequence of $N$-bit vectors. For example, one might consider the case
where the different binary words carry information about the system at
increasing temporal depths, e.g. by giving the $N$-bit description of
the system at every past tick of a clock. In general, the {\it state}
of the system will be given by
	$$ S = \{ {\bf S} \left( 0 \right), {\bf S} \left( 1
	\right) , \cdots , {\bf S} \left( n \right) , \cdots \}, $$
where $ {\bf S} \left( n \right) $ is a vector with components $
S_i\left( n \right) = 0, 1 $, 
($i = 1, \cdots, N $). The set of such binary states will be denoted 
by $ \Omega $.

	The approximation which makes this concept practical, akin to
the $ 128$-bit version of floating-point variables, is the
truncation of the symbolic states to the first $ n $ words.
This is a good approximation if the difference between states which
coincide in the first $ n $ words belongs to a small neighborhood of
the origin. We will formalize this demand through the assignment of a
base for the topology on $ \Omega $, related to the cylinders of
symbolic dynamics~$ {}^{[10, 13]} $.   

	With this topology, the space $ \Omega $ is homeomorphic to
the one-sided shift space of symbolic dynamics. Our main contribution
is to provide a definition of chaos for general dynamical maps in
$\Omega$. In symbolic dynamics one usually considers the shift map
$\sigma$, which consists in erasing the word $ {\bf S}(0) $ from the
semi-infinite sequence and shifting the other slices by ${\bf S}(n)
\longrightarrow {\bf S}(n - 1)$~$ {}^{[10]} $.  This example satisfies
our definition of discrete chaos. But we stress that this is only one
of many possible chaotic maps in $\Omega$.

	We will consider functions which are continuous or
discontinuous. Neural networks and cellular automata will turn out to
be examples of discontinuous functions. For general discontinuous
functions very little is known, basically due to the fact that
analytically there is very little that one can prove. However,
numerically one can distinguish several types of dynamical behavior. In $
\IR^N $ the Grassberger and Procaccia method is widely used to estimate 
the fractal dimension of attractors. We will extend its application
to the space $ \Omega $ in order to characterize different chaotic
behaviors and define an effective attractor dimension. 

 One important class of maps which we will consider in 
this paper corresponds to the case when the binary state represents 
the system at every past tick of a clock, as explained above. To define 
such a map one must provide a function which allows one
to compute the new word $ {\bf S}(0) $ from the state $ S $. 
The left inverse of any such map is the shift map $ \sigma $ of symbolic 
dynamics. Examples include neural network models 
and cellular automata. Note that in this case not all points of $ \Omega $
represent possible histories: Instead, $ \Omega $ plays the role of
an embedding space for the attractor.

	The results of this paper can be generalized without
difficulty to other alphabets besides the binary one, and also to the
case where the space $\Omega$ is the two-sided shift space~$ {}^{[11,
12]} $, where a state is given by a sequence
	$$ \left\{ \cdots {\bf S} \left( - 1
	\right), {\bf S} \left( 0 \right), {\bf S} \left( 1 \right), \cdots
	\right\}. $$ 

	In this case our construction reduces to the invertible shift
map when ${\bf S}(n)$ is taken to be the binary description of the
system but once again we stress that this is only one of several
possible dynamical maps $F: \Omega \longrightarrow \Omega$.

	The organization of this paper is as follows. ``Discrete
chaos'' will be defined in Sec.~2 and different types of dynamical
maps in $ \Omega $ are discussed. In Sec.~3 we will consider the
correlation function $C(\rho)$ and the Grassberger Procaccia method to
compute the correlation dimension of the attractor.  Numerical
examples will be considered in Sec.~4 . In Sec.~5 we give the
conclusion.

\

\section{2. Chaotic Dynamics of Binary Systems.}

	Binary systems, like cellular automata and neural networks, are
described, in general, by a set of $ N $ binary variables $ S_i $ $ i
= 1, \dots , N $, or in short $ {\bf S} $, that evolve according to
dynamical rules. The natural metric for these systems is the
Hamming distance
	$$ d_H \left( S - S' \right) \equiv \sum_{i = 1}^N \mid S_i -
	S'_i\mid . $$ 

The space $ \left\{ {\bf S} \right\} $ has $ 2^N $ possible states and
so the topology constructed from $ d_H $ is
discrete. Generally one is interested in studying these dynamical
systems in the limit $ N \to \infty $ since that is where interesting 
statistical properties appear, such as
phase transitions, and it is possible to use powerful
techniques like mean field theory~${}^{[1-4]}$. Furthermore
numerical simulations which need to be done for finite, but
large $ N $, are understood as approximations of a system with
infinite variables, much in the same way as floating
point variables in computers are finite approximations of
real numbers which generally have an infinite number of
digits. Nevertheless for $ N \to \infty $, $ d_H $ is no longer 
a distance and the topology is ill defined in that
limit. That makes our understanding of binary systems
quite different from that of dynamical systems in $
\IR^d $ or in differentiable manifolds where one works with the usual 
topology of the real numbers. Here we will overcome this situation
by extending the phase space $ \left\{ {\bf S} \right\} $ to have an
infinite number of states while preserving the equal status that the
Hamming distance confers to each of the variables. That is to say, all
the variables $ S_i $ give the same contribution to the distance for
any $ i $.

	Let us consider the Cartesian product of infinite copies of $
\left\{ {\bf S} \right\} $ and call this space $ \Omega $. We denote
the elements of $ \Omega $ by 
	$$  S = \left({\bf S} \left( 0 	\right), {\bf S} \left( 1
	\right), {\bf S} \left( 2 	\right),... \right) . 	\eqno(1)
	$$

	We make $ \Omega $ a topological space by introducing the
following base:
	$$ {\cal N}_{n} \left( S \right)  = \left\{ S' \in \Omega
	\vert {\bf S'} \left( m \right) = {\bf S} \left( m
	\right), \forall m < n \right\}, \eqno(2) $$
with $ n = 1, 2, \dots $. These base sets are closely related to 
the cylinders in one-sided shift spaces and $
\Omega $ is homeomorphic to the space of symbols of 
the symbolic dynamics with $2^N $ symbols~${}^{[10, 11]}$. It follows 
that $ \Omega $ is a cantor set. In symbolic
dynamics the topology is usually derived from the metric
	$$ d \left( S, S' \right) = \sum_{n = 0}^{\infty} {1
	\over2^n}\ d_n \left( S - S' \right), \eqno(3) $$ 
where
	$$ d_n \left( S - S' \right) \equiv \sum_{i = 1}^N \mid
	S_i(n) - S'_i(n) \mid . \eqno(4) $$
is the Hamming distance of the $ n^{th} $ copy of $ \left\{ {\bf S}
 \right\} $. One can check that if $ {\bf S} \left( m \right) =
{\bf S'} \left( m \right) $ $ \forall m < n $ then $ d \left( S, S'
\right) < {N + 1 \over 2^{n - 1}} $, so that (2) and (3) define the
same topology.

	 Here and in the following our purpose is to study dynamical
systems in $ \Omega $ generated by a function 
$ F: \Omega \longrightarrow \Omega $. This function may be continuous
or discontinuous, unless explicitly stated below. 
Allowing discontinuous functions in principle opens the door to a richer 
variety of systems, which include neural networks and cellular automata.

	We begin by generalizing in a natural way the definitions of
chaos in subsets of $ \IR^N $ (see for example Ref.~[11]) to $\Omega
$.

\noindent {\it Definition 1}: $ F $ has {\it sensitive dependence on
initial conditions} on $ {\cal A} \subset \Omega $ if $ \exists n \in
\IN $ $ \ni \forall S \in {\cal A} $ and $ \forall {\cal N}_m \left(
S \right) $ $ \exists S' \in {\cal N}_m \left( S \right) \cap {\cal
A} $ and $ k \in \IN $ such that $ F^k \left( S' \right) \notin {\cal
N}_n \left( F^k \left( S \right) \right) $.

\noindent {\it Definition 2}: Let $ {\cal A} \subset \Omega $ be a
closed invariant set. $ F: \Omega \longrightarrow \Omega $ is {\it
topologically transitive} on $ {\cal A} \subset \Omega $ if for any
open sets $ U, V \subset {\cal A} $ $ \exists n \in \IZ $ $ \ni F^n
\left( U \right) \cap V \not= \emptyset $. In the last expression, if
$ F $ is non invertible we understand $ F^{-k}(U) $ with $ k > 0 $, as
the set of all points $ S \in \Omega $ such that $ F^k(S) \in U $.

\noindent {\it Definition 3}: Let $ {\cal A} \subset \Omega $ be a
compact set. $ F: {\cal A \longrightarrow A} $ is {\it chaotic} on $
{\cal A} $ if $ F $ has sensitive dependence on initial conditions
and is topologically transitive on $ {\cal A} $.

\noindent {\it Definition 4}: A closed subset $ {\cal M} \subset
\Omega $ is called a {\it trapping region} if $ F \left( {\cal M}
\right) \subset {\cal M} $.

\noindent {\it Property 1}: If $ F $ is a continuous function in $
\Omega $, $ F^n \left( {\cal M} \right)$ is compact
and closed $ \forall n \in \IN $.

\noindent {\it Proof}: Since every closed subset of a compact set is
compact, it follows that $ {\cal M} $ is compact and since $ F $ is
continuous $ F^n \left( {\cal M} \right) $ is compact. Since $ \Omega
$ is Hausdorff every compact subset of it is closed, so $ F^n \left(
{\cal M} \right) $ is closed~${}^{[14]}$.  \hfill \qed

\noindent {\it Definition 5}: The map $ F: \Omega \longrightarrow
\Omega $ has an {\it attractor} if it admits an asymptotically
stable transitive set, i.e., if there
exists a trapping region $ {\cal M} $ such that 
	$$ \Lambda \equiv \bigcap\limits_{n \geq 0} F^n \left( {\cal
	M} \right) $$ 
and $ F $ is topologically transitive on $ \Lambda $.

Note carefully that the trapping region is defined in the $ \Omega $
space while in the theory of dynamical systems in manifolds, it is
defined in the manifold~${}^{[10-13, 15]}$. This makes, most theorems
(as those shoed in Ref.~[15]) concerned with Cantor sets considered as
attractors in manifolds to be not applicable.

\noindent {\it Property 2}: If $ F $ is a continuous function in $
\Omega $, $ \Lambda $ is compact and closed.

\noindent {\it Proof}: From property~1 if $ F $ is continuous, 
$ \Lambda $ is an intersection of closed sets, so it is closed. Since
every closed subset of a compact space $ \Omega $ is compact, it
follows that $ \Lambda $ is compact.  \hfill \qed 

\noindent {\it Definition 6}: $ \Lambda $ is called a {\it chaotic
attractor} if $ F $ is chaotic on $ \Lambda $.

\noindent {\it  Lemma}: Let $ F $ be a continuous function in $
\Omega $, if $ \Lambda $ is a chaotic attractor then it is perfect.

\noindent {\it Proof}: By property 2, $ \Lambda $ is closed, it
remains to prove that every point in $ \Lambda $ is an accumulation
point of $ \Lambda $.  By contradiction, let $ S_0 \in \Lambda $ be an
isolated point, then there exists $ n \in \IN $ $ \ni $ $ {\cal N}_n
\left( S_0 \right) \cap \Lambda = \left\{ S_0 \right\} $. Then, by
topological transitivity $ \Lambda $ has an isolated orbit (the orbit
of $ S_0 $) which implies that it is not sensitive to initial
conditions on $ \Lambda $.
\hfill \qed

\noindent {\it Theorem}: If $ F $ is a continuous function in $
\Omega $, and $ \Lambda $ is a chaotic attractor then it is a Cantor set.

\noindent {\it Proof}: The theorem follows directly from property 2,
the Lemma and the fact that a subset of a totally disconnected set is
also totally disconnected. \hfill \qed

	In the following we will consider some
examples of dynamical functions $ f : \Omega \longrightarrow \Omega $.
The first one is the one-side shift map $ \sigma $ of
symbolic dynamics which we introduce to familiarize the reader with
the notation.

\noindent {\it i}) The one-sided shift map $ \sigma $.

	The continuous map $ \sigma $ defined by
	$$ \sigma \left( {\bf S} \left( 0 \right), {\bf S} \left( 1
	\right),... \right) = \left( {\bf S} \left( 1 \right), {\bf
	S} \left( 2 \right),... \right), \eqno(5) $$ 
is chaotic in $\Omega$~${}^{[10]}$.
Note that $ \sigma $ is non-invertible and its action
loses the information carried by the binary state $ {\bf S}
\left( 0 \right) $. The meaning and usefulness of this map is quite clear
in the context of symbolic dynamics when the Conley-Moser conditions
are satisfied~${}^{[16]}$. There one studies, in general, a
non-invertible function $ f : \Xi \longrightarrow \Xi $ where $ \Xi $
is a Cantor set embedded in $ \IR^N $. The set $ \Xi $ is divided in
$ 2^N $ sectors $ I_\alpha $ $ \alpha = 0, 1,..., 2^N $.  Then it is
possible to establish a topological conjugation between $ f $ and $
\sigma $ through a homeomorphism $ \psi $, so that the following
diagram commutes~${}^{[11]}$ 

\baselineskip=12pt
	$$ \eqalign{&\Xi \buildrel f \over \longrightarrow \Xi \cr
	{}_\psi & \downarrow  \ \ \quad \downarrow {}_\psi \cr &
	\Omega \buildrel \sigma \over \longrightarrow \Omega \cr}.
	\eqno(6) $$ 
Moreover, let $ S = \psi \left( x \right) $, then $ {\bf S} \left( 
n \right) $ is the binary decomposition of the label $ \alpha $, such 
that $ f^n \left( x \right) \in I_\alpha $.

\noindent {\it ii}) Chaotic maps with non-trivial attractors 
in $\Omega$.

	The shift map can be modified to create maps which are
homeomorphic to the shift map on an asymptotically stable transitive
subset of the space of symbols. We introduce two very simple examples:

	Take the space of symbols $ \Omega $ with $ N = 2 $, 
homeomorphic to $ \Xi \times \Xi $ where $ \Xi $ is the space of
symbols with $ N = 1 $, that is the space of semi-infinite sequences $
S = \left( S_0, S_1, S_2,... \right) $. Then consider the function $
f_c : \Xi \times \Xi
\rightarrow \Xi \times \Xi $ given by $ f_c = \sigma \times \zeta
$. Where $ \sigma $ is the usual shift function and $
\zeta $ is a right inverse of the shift function defined as follows:
	$$ \zeta \left( S_0, S_1, S_2,... \right) = \left( 0, S_0,
	S_1, S_2,... \right). $$
It is easy to check that $ \zeta $ is a continuous function, and of 
course so is the shift: so  $f_c $ is continuous. The set 
$ \Xi \times \left\{ 0 \right\} $ is an asymptotically stable 
transitive set, on which the restriction of $ f_c $ is the shift map
$ \sigma $.

	As another example, consider the space $ \Omega $ with $ N = 1 $. 
It can be split into the disjoint union of two Cantor sets 
$ \Omega = \Lambda_0 \cup \Lambda_1 $. Where $ \Lambda_0 $ is the set 
of sequences such that $ S_0 = 0 $ and an analogous fashion for 
$ \Lambda_1 $. Take the continuous function 
$ f_\pi = \pi \circ \sigma $, where $ \sigma $ is
the shift map and $ \pi $ projects $ \Omega $ in $ \Lambda_0 $ such
that:
	$$ \pi \left( S_0, S_1, S_2, ... \right) = \left( 0, S_1, S_2,
	... \right). $$ 
Then the action of $ f_\pi $ is given by,
	$$ f_\pi \left( S_0, S_1, S_2, ... \right) = \left( 0, S_2,
	S_3, ... \right). $$
It is easy to check that $ \Lambda_0 $ is a chaotic attractor of $
f_\pi $.

\noindent {\it iii}) Chaotic maps in $ \Omega $ induced through
chaotic maps in Cantor subsets of $ \IR^N $.

	We will consider a homeomorphism which relates a Cantor set $
\chi \subset \IR^N $ to the space $ \Omega $ and allows one to
construct chaotic maps in $
\Omega $ from chaotic maps in $ \chi $ through topological
conjugation. Let $ \chi \subset \IR^N $ be the Cantor set that results
from taking the Cartesian product of $ N $ Cantor sets $ \chi_i$;
	$$ \chi = \bigotimes\limits_{i = 1}^N \ \chi_i, $$
where the $ i^{th} $ component $ \chi_i $ is constructed by
suppressing from the interval $ \left[ 0, 1 \right] $ the open middle
$ 1 / a_i $ part, $ i = 1, \dots, N $, $ a_i > 1 $, and repeating this
procedure iteratively with the sub-intervals, see Fig.~1. Now,
we define $ \phi: \Omega \longrightarrow \chi $ by:
	$$ \phi_i \left( S \right) = \sum_{n = 1}^\infty \left( l_{n
	- 1} - l_n \right) S_i \left( n - 1 \right) \eqno(7) $$ 
where
	$$ l_n = {1 \over 2^n} \left( 1 - {1 \over a_i} \right)^n
	\eqno(8) $$ 
is the length of each of the remaining $ 2^n $ intervals at the $
n^{th} $ step of the construction of $ \chi_i $. If $ \Omega $ is
endowed with the metric (3) and $ \chi \subset \IR^N $ with the
standard Euclidean metric, is easy to show that $ \phi $ is a
homeomorphism.

	Now, if we have a map $ f: \IR^N \longrightarrow \IR^N $
which is chaotic in $ \chi $ we can construct a map $ F: \Omega
\longrightarrow \Omega $ which is chaotic in $ \Omega $, and is
defined through the commutation of the diagram

\baselineskip=12pt
	$$ \eqalign{&\chi \buildrel f \over \longrightarrow \chi
	\cr {}_\phi & \uparrow \ \ \quad \uparrow {}_\phi \cr &
	\Omega \buildrel F \over \longrightarrow \Omega \cr}.
	\eqno(9) $$ 
This leads to an interesting practical application of the
homeomorphism $ \phi $, to realize computer simulations of chaotic
systems on Cantor sets. If, for example, one iterates the logistic
map $ f \left( x \right) = \mu x \left( 1 - x \right) $ 
 for $\mu \geq 4$ with a floating-point variable,
the truncation errors nudge the trajectory away from the Cantor set and
eventually $x \to -\infty$. The homeomorphism $\phi$ suggests a natural
solution to this, which is to iterate the truncated binary states
rather than the floating-point variable. To iterate the dynamics, one
computes $ x_i = \phi_i (S) $ $ \forall i = 1, \dots , N $ by
assuming that the truncated bits are all equal to zero, then applies
$ f $ to obtain $ x' = f(x) $. Since $ x' $ generally does {\it not}
belong to the Cantor set (because of truncation errors), in the
process of constructing $ S' = \phi^{-1}(x')$, at some $ n $ one will
find that this point does not belong to either the interval
corresponding to $ S_i (n) = 0 $ or to $ S_i (n) = 1 $. This
truncation error can be corrected by moving to the extremity of the
interval which lies closest to $ x'_i $. In this way, truncation
errors are not allowed to draw the trajectory away from the Cantor
set $ \chi \subset \IR^N $.  

\noindent {\it iv}) Binary systems with memory.

	Now we are going to define a map $ \Gamma : \Omega
\longrightarrow \Omega $ which is very useful to analyze binary
systems with causal deterministic dynamics on $ N $ bits,
such as neural networks, cellular automata, and neural networks with
memory~${}^{[1-4, 17]}$. Let 
	$$ \gamma_i : \Omega \longrightarrow \left\{ 0, 1 \right\},
	\eqno(10) $$
$ i = 1,...,N $, be a set of  continuous or discontinuous 
functions. $ \Gamma : \Omega \longrightarrow \Omega $ is then defined by:
	$$ \Gamma_i \left( S \right) = \left( \gamma_i \left(
	S \right), S_i \left( 0 \right), S_i \left( 1 \right),
	\dots \right).  $$ 
or in a short hand notation
	$$ \Gamma \left( S \right) = \left( \gamma \left(
	S \right), S \right).  \eqno(11) $$ 
Such maps have the following properties.

\noindent {\it Property 3}  The shift
map (5) is a left inverse of $ \Gamma $ since from (11) $ \sigma
\circ \Gamma \left( S \right) = S $. If $ \Omega $ has an attracting
set $
\Lambda \subset \Omega $, then $ \sigma $ is also a right inverse in
the restriction of $ \Gamma $ to $ \Lambda $, so that, $ \Gamma
\mid_\Lambda^{- 1} = \sigma $. 

\noindent {\it Proof}: $ \forall S \in
\Lambda $ $ \exists S' \in \Lambda $ such that $ \Gamma \left( S'
\right) = S $. Since 
	$$ \Gamma \left( S' \right) = \left( \gamma \left( S'
	\right), S' \right) = S$$
and
$$S = \left( {\bf S} (0), S_1 \right) ,$$  
where $ S_1 \equiv \left( {\bf S} (1), {\bf S} (2), \dots \right)
$, one sees that $ S' = S_1 $. Thus,
	$$ \Gamma \circ \sigma \left( S \right) = \Gamma \left( S_1
	\right) = \Gamma \left( S' \right)  = S. $$ \hfill \qed

\noindent {\it Property 4} $\Gamma$ has an attracting set $ \Lambda $ contained
properly in $ \Omega $.

\noindent {\it Proof}: Given $S$ there are $2^N$ states 
$S' = ({\bf S'}(0), S)$ of which only one, $\Gamma(S) = (\gamma(S),
S)$, belongs to $\Gamma(\Omega)$. Therefore the set
	$$ \Lambda \equiv \bigcap\limits_{n \geq 0} \Gamma^n 
	\left( {\Omega} \right) $$ 
is a proper subset of $\Omega$. \qed

\noindent {\it Property 5} If $\Gamma$ is continuous, then it is not
sensitive to initial conditions.

\noindent {\it Proof}: $\Gamma$ is a continuous map on a compact set, so 
it is uniformly continuous. Therefore there exists a $\delta > 0$ such
that for any $S \in \Omega$, $d(S', S) < \delta \Rightarrow
\gamma(S) = \gamma(S')$ and hence $d(\Gamma(S), \Gamma(S')) < \delta /
2$, where the distance function is given by (3). 
Applying the same argument to each iterate $\Gamma^k(S)$ shows that
$d(\Gamma^k(S), \Gamma^k(S')) < \delta / 2^k$, which contradicts sensitivity 
to initial conditions. \qed

\noindent {\it Property 6} If $ \Gamma $ is continuous, then the
attractor $\Lambda$ is finite. 

\noindent {\it Proof}: From Property 4 we know that $\Lambda$
exists. The property then follows from Property 5 above: Indeed, if
$\Gamma$ is not sensitive to initial conditions, then there is a $n >
0$ such that $ \forall S \in \Omega $ 
	$$ \lim\limits_{k \to \infty} \ d \left( \Gamma^k (S) - \Gamma^k (S')
	\right) = 0 $$ 
$ \forall S' \in {\cal N}_n (S) $.  The set $A
\subset \Omega$ defined by $S \in A $ iff
$\forall m > n, {\bf S}(m) = 0$, has a finite number of elements,
namely $2^{N \times n}$. The whole space $\Omega$
is the union of the $n-$neighborhoods of each element of $A$, and as
we just showed the map $\Gamma$ is contracting in each such neighborhood,
so the number of points in the attractor cannot be greater than 
the number of elements of $A$, namely $2^{N \times n}$. \qed

	Neural networks and cellular automata are binary dynamical
systems in which the values of the state variables $ S_i $, $ i = 1,
\dots, N $, at time $ t $ depend on the state variables at time $ t -
1 $. These systems are described by a function $ \Gamma $ such that
the functions $ \gamma_i $ depend only on the components $ {\bf S}
\left( 0 \right) $. Therefore, all points $S' \in {\cal N}_n(S)$ for
$n > 0$ have the same evolution so that these systems are not
sensitive to initial conditions. One can recover a very rough
approximation of sensitive dependence on initial conditions by
considering the growth of {\it Hamming distance} with time, rather
than the metric (3) of symbolic dynamics. However, one cannot describe
the behavior of these systems to be approximately chaotic: They are
well known to have attractors that consist of a collection of periodic
limit-cycles, and as we will see in Sec.~4, the points of these
limit-cycles are scattered over configuration space without any
effective lower-dimensional structure. In particular, given any one
point on the attractor there is usually no other point ``nearby'',
even in the weak sense of the Hamming distance, that also belongs to
the attractor. This fact makes most practical uses of chaos theory in
prediction and control inapplicable.

\noindent {\it v}) A compact topology for neural networks and
cellular automata.

	Since neural networks and cellular automata in general are
systems in which all the variables have the same type of
interactions, it is natural to consider the Hamming distance as the
metric (it is in fact the most widely used metric in the literature, see for
instance Ref.~[1-4] and the references therein). We have already seen
that the topological structure which the Hamming distance confers to
the phase space does not conduce to chaotic behavior in the sense that
we understand it even if we extend the phase space to $ \Omega
$. However, not all the neural network and cellular automata models
confer the same type of interactions to neurons, so the use
the Hamming distance for the metric is not so compelling. The use of
a different metric can lead to a completely different topology.
The resulting system will in general display a very different 
dynamical behavior. For example the map $ x_{n + 1} = \alpha x_n $ 
produces quite different dynamical behaviors for $x_n \in \IR $ and 
$ x_n \in S^1 $.

	So, let us consider systems which evolve according to the rule
	$$ R_i \left( t + 1 \right) = f_i \left( R \left( t \right)
	 \right) \eqno(12) $$
$ R_i = 0, 1 $; $ i = 1,..., M $ and take for the metric 
	$$ d \left( S, S' \right) = \sum_{n = 0}^M {1 \over2^n}\ d_n
	\left( S - S' \right). \eqno(13) $$ 
These systems include neural networks and cellular automata as
particular examples, but where the weight of the different neurons
drops off as $2^{-n}$. The metric (13) remains well defined in
the limit $ M \to \infty $ and once again we obtain 
the space $ \Omega $. In fact (12) and (13) with $ M \to
\infty $ are equivalent to (3) and (4) with $ N = 1 $ and $
S_1 \left( n \right) = R_n $. As we will see in the next section
these systems can have a correlation dimension which is less than or equal
to one. 

\

\section{3. Correlation Function.}

	In the theory of dynamical systems in $ \IR^N $ one is 
interested in calculating the fractal dimension of
the attractor in which the system evolves. To do so, following the
method of Grassberger and Procaccia~${}^{[18]}$ one defines the
correlation function $ {\cal C} \left( \rho \right) $ as the
average of the number of neighbors $ S_t $, $ S_{t'} $, with $ S_t = 
 F^t\left( S \right) $, which
have a distance smaller than $ \rho $. Since in $ \IR^N $
the volume of a sphere of radius $ \rho $ grows like $ \rho^N $, one
identifies the correlation dimension $ D_a $ of the attractor with the 
growth rate in $ {\cal C} \left( \rho \right) \sim \rho^{D_a} $. This
leads to the definition of the correlation dimension as 
	$$ D_a = \lim\limits_{\rho, \rho' \to 0} \ {\left( \log
	\left( {\cal C} \left( \rho \right) \right) - \log \left(
	{\cal C} \left( \rho' \right) \right) \over \log \left( \rho
	\right) - \log \left( \rho' \right) \right)}. \eqno(14) $$ 
In order to have an analogous methodology to compute correlation
dimensions in $ \Omega $, it is necessary to know how many states $ S'
$ are within a distance less than $ \rho $ from a given point $ S $.
Since $ \Omega $ is homogeneous we can take $ S = 0 $. To do the
calculation we make $ \Omega $ into a finite space by truncating the
semi-infinite sequence to only $ T $ slices, and take the limit $ T \to
\infty $ in the end, that is:
	$$ {\cal C} \left( \rho \right) = \lim\limits_{T \to \infty} 
	{1 \over
	2^{N	T}}\sum_{\left\{ S \right\}}\Theta \left( \rho - d \left( S, 0
	\right) \right), \eqno(15) $$ 
where the distance is given by (3). Expressing $ \Theta ( x ) $ in
terms of its Fourier transform $ \omega \left( k \right) = \pi \delta
\left( k \right) - {i \over k} $ we have
	$$ {\cal C} \left( \rho \right) = \lim\limits_{T \to \infty} 
	{1 \over
2^{N 	T}} {1 \over 2 \pi} \int_{- \infty}^{+ \infty} dk \ \omega
\left( k \right) 	e^{i k \rho} \sum_{\left\{ S \right\}} {\rm e}^{- i
k d \left( S, 	0 \right) }. $$ 
The sum over $ \left\{ S \right\} $ can be evaluated easily obtaining 
	$$ \sum_{\left\{ S \right\}}  e^{- i k d \left( S, 0 \right) } =
	2^{N T}  e^{- i k N} \left( \prod \limits_{n = 0}^T \cos {k \over
2^{n + 1}} \right)^N. $$ 
Using the identity $ \sin k / k = \prod_{n = 0}^\infty \cos {k \over
2^{n + 1}} $ we obtain the integral 
	$$ {\cal C} \left( \rho \right) = {1 \over 2 \pi} \int_{-
\infty}^{+ \infty} dk \ \omega  \left( k \right) \left( {\sin k \over
k} \right)^N {\rm e}^{i 	k \left( \rho- N \right) }, $$
which may be evaluated by standard complex variable methods, to obtain
the final result for the correlation function in $ \Omega $,
	$$ {\cal C} \left( \rho \right) = {1 \over 2^N N!} \sum_{k = 0}^{
\left[ \rho / 2 \right]} \left( - 1 \right)^k \pmatrix{
	N\cr k\cr} \left( \rho - 2 k\right)^N. \eqno(16) $$
So we see that the scaling in $ \Omega $ is not a power law as in 
$ \IR^N $. However in the definition of the attractor dimension one
is interested in calculating $ {\cal C} \left( \rho \right) $ for $
\rho \to 0 $. For $ \rho \leq 2 $ equation (16) has the form 
	$$ {\cal C} \left( \rho \right) = {1 \over 2^N N!} \rho^N
	.\eqno(17) $$
Therefore, the same techniques applied in $ \IR^N $ can be used in $
\Omega $, in particular an effective ``attractor dimension'' will 
be given by (14). 

\
 
\section{4. Numerical Examples.}

	We have examined numerically several of the binary systems
which have been considered in the literature, including the random $
k=4 $ cellular automata~${}^{[7]}$, and some neural network models
such as those studied by Crisanti {\it et. al.} in the context of
Shannon's entropy~${}^{[2]}$. 

	Random cellular automata of rank $ k $ consist of $ N $ binary
variables $ S_i = 0, 1 $ with the following dynamical rule.
For each binary variable one chooses at random a boolean function 
$ f_i : \IZ_2^k \longrightarrow \IZ_2 $ from  $ k $ binary variables 
into one, among the $ 2^{2^k} $ possible functions. Then, for 
each $ i = 1, \dots , N$, a set of $ k $ numbers 
$ \left\{ i_1, i_2, \dots , i_k \right\} $
is selected at random from $ \left\{ 1, \dots , N \right\} $. These 
numbers are interpreted as labels for the $k$ inputs of the boolean
function at that $i$. The evolution of the system in time is given by
applying the boolean rules synchronously at the $N$ variables:
	$$ S_i \left( t + 1 \right) = f_i \left( S_{i_1} \left( t
	\right), \dots, S_{i_k} \left( t \right) \right). \eqno(18) $$ 
For $ k \geq 4 $ the system has very long cycles with periods of
order $ e^N $ and present the phenomenon of damage spreading which is
the standard way in which a first ``Lyapunov exponent'' is assigned
to such systems.  

	In Ref.~[2] Crisanti {\it et. al.} studied a binary neural network
described by variables $ S_i = \pm 1 $. The variables evolve in parallel
according to the rule
	$$ S_i \left( t + 1 \right) = {\rm sgn} \left( \sum_{j = 1}^N J_{ij}
	S_j \left( t \right) \right),\eqno(19) $$ 
where 
	$$ J_{ij} = J_{ij}^S + k J_{ij}^A $$
is the synaptic matrix, with $ J_{ji}^S = J_{ij}^S $ and $ J_{ji}^A =
 - J_{ij}^A $ being random independent gaussian variables with mean
zero and variance $ \sigma^2 = 1 / (N - 1)(1 + k) $. The parameter $
k $ measures the amount of asymmetry of the synapses. For $ k = 0 $
the matrix is symmetric and the network has fixed points
as attractors. For $ k > 0 $ it is asymmetric and the
network can have limit cycles as attractors. When $k > k_c = 0.5$, long
limit cycles are obtained, with period of
order $ e^N $ but with fluctuations of the same order. In
Ref.~[2] the Shannon entropy has been calculated numerically in the
range $ k_c < k < 0.9 $ and the scaling was found to be given by
	$$ h \sim \left( k - k_c \right)^{1 / 2}. $$ 
In the limit $ k \to 1 $, $h$ attains a value which is very close to
the maximum value $ \log (2) $, characteristic of a random process. All
of this indicates a high degree of complexity.

	However, for both of the dynamical systems (18) and (19), the
dynamics is not topologically transitive on the limit set of all
periodic orbits: As we will see below, points on this set are isolated
in Hamming distance, so most points do not even have any
``near-neighbors'' that might attempt to satisfy the conditions in
the definition of topological transitivity.
 
	The number of returns to within a Hamming distance $ d_H $ of
an initial point on one of the long periodic orbits is given in the
[Figs.~2, 3] for the cellular automata (18) with $ k = 4 $ and the
neural network (19) with $ k = 1.2 $. We have also graphed the
best-fitting Gaussian curve for comparison. The first obvious result
is that there are no returns with $d_H < 50$ in an automata with $ N =
200 $, in $ 5 \times 10^6 $ iterations of the dynamical map. The
neural network was run with $N = 100$ neurons for $ 5 \times 10^5 $
iterations, and again no returns were found with $d_H < {N \over
4}$.  Both the value of the nearest return and the fit to a Gaussian
are consistent with a random process which produces patterns all over
the configuration space without any restriction to a possible
``attracting subspace''. This indicates a very high degree of
algorithmic complexity~${}^{[8]}$ in the time-series, which
reflects a lack of predictability, like Shannon entropy which is a
statistical measure of disorder.  

	Without anything analogous to a transitive ``attractor'', 
none of the practical applications of chaos theory can
carry through for large values of $N$.  The phase space 
reconstruction methods and other
versions of the ``method of analogues''~${}^{[19]}$ fail because {\it
one finds no good analogue} in any finite data set, for large $N$.
The lack of close returns in binary systems can often be related to the 
failure to find an attractor on which the dynamics is topologically
transitive. 

	Another example, more in the spirit of the maps $\Gamma$ is an
asymmetric neural network with state-dependent synapses originally
designed to recognize sequences of patterns, and described in
Ref.~[4]. As shown there, this system has a transition from a stable
sequence reproduction to a disordered behavior.  We shall modify the
dynamical rule by introducing a memory in an analogous way as has
been done in Ref.~[17], as follows: 

	$$S_i(t+1) = {\rm sgn}\left( \sum_{n=0}^{T - 1} \ {1 \over 2^n}
	\sum_{j = 1}^N\ J_{ij}^{(n)} S_j(t-n) \right), \eqno(20) $$ 
where the synapses is given by 
	$$J_{ij}^{(n)} = {1 \over N} \sum_{\mu=1}^p\ \xi_i^{\mu+n+1} 
	\xi_j^{\mu}, \eqno(21) $$
and 
	$$S_{\mu} = {1 \over N} \sum_{i=1}^N\ S_i \xi_i^{\mu}$$
is the correlation of the state of the network with the pattern $
\xi^\mu $. The patterns $ {\bf
\xi}^{\mu} = \pm 1 $ with $ \mu = 1, \dots , p + T $ are random
independent, equiprobable variables, and $ p $ is a parameter of the
model. The reader not familiar with the notation of Hopfield-type
neural networks may refer to Ref.~[4].

	It is easy to show that the argument of the sign function
referred to above as function $\gamma_i$, does in fact vanish in
$\Omega$ when $T \to \infty$. So the map $\Gamma: \Omega \to \Omega$
is discontinuous, and one cannot immediately rule out the possibility
of non-periodic orbits~${}^{[20]}$.  In practice however one always
uses a finite memory in computer simulations, so the continuity is
recovered and it is the sensitivity to initial conditions which is not
valid.

	In applications it is often the case that two points can be
considered to be distinguishable only if their mutual distance is
greater than a small ``cutoff'' value $\lambda$. If $\lambda > 1/2^T$
one can then claim that the map is ``effectively'' sensitive to
initial conditions, which suggests that the ensuing dynamics may be
correctly described as being ``approximately chaotic''. This is best
analyzed by computer simulation.

 	We have run this system with $T = 30$, $ N = 49 $ neurons and
$ p = 19 $ and find evidence for a non-trivial attractor with a low
effective dimension: Unlike in the examples above there are
substantially more near-returns than for a random sequence, as shown
in [Fig.~4], and the correlation function $ {\cal C} ( \rho ) $
produces an effective dimension $ D_a \approx 2 $ in the range $ 0.002
< \rho < 0.02 $ (see Fig.~5).

	For both the neural network (19) and the $ k=4 $ cellular
automata (18), we observe a very different behavior as expected; the
correlation graph $ {\cal C} (\rho) $ coincides for all $ \rho $, with
(16), within an accuracy of the $ 98 \% $ which means that $ D_a
\approx N $. That means that the orbits are scattered in the whole
space $ \Omega $ as one would expect from the fact that the
distribution of returns to Hamming distance $ d_H $ is approximately
Gaussian.

\

\section{5. Conclusion.}

	From an initial {\it ansatz}, to replace the usual
idealization of physical states as ``points'' on a differentiable
manifold by another idealization as infinite ``binary states'', we
proceeded to define a topology which makes the truncation to finite
states a valid approximation, in the same sense that the usual
topology on $\IR^N$ allows one to approximate a real coordinate by a
finite string of digits or bits. This lead us to a space $ \Omega $
which is homeomorphic to the space of symbolic dynamics.

	Continuous or discontinuous dynamical maps on the space of
symbolic states can lead to attracting sets within $ \Omega $, in
which case an attractor is defined in the usual way. The dynamical map
is said to be chaotic on the attractor if it is sensitive to initial
conditions and topologically transitive.

	 Finite systems such as neural networks and cellular automata
without memory that depend only on the previous time step and for
which the different bits have comparable importance do {\it not}
provide good approximations of chaos when the Hamming distance is used
as the metric . The absence of even an approximate manifestation of
chaos has important practical consequences - for example we found that
prediction models based on a search for analogous examples in a data
set are not applicable because no good analogues are found in any
reasonable amount of data.

	The practical value of the analysis of dynamical maps on the
space $\Omega$ is probably limited to special complex systems problems
where an extended binary description is more natural than a continuum
description. Discrete space-time formulations of quantum gravity offer
another potentially rewarding area of applicability: There, the
evidence for discrete small-scale structure combine with the perceived
need of a space-time ``sum over histories'' interpretation lead to a
formalism where one defines ``states'' to be truncated discrete
space-time histories. An interesting example is the causal set
formalism, where a partially ordered set or Poset is conjectured to
constitute the minimal required structure to formulate a theory of 
quantum gravity.

A priority in the continuation of this work is to further elucidate
the chaotic properties of neural networks and cellular automata when
a compact metric compatible with the topology of $ \Omega $ is given. 

\

\

\noindent {\bf Acknowledgments} 

\noindent The authors would like to thank C. Stephens and M.
Socolovsky for helpful comments and discussions; V. Dom\'\i nguez,
E. Sacrist\'an, F. Toledo for computational advice. The first author
(HW) would also like to thank Th. W. Ruijgrok for helpful
conversations at Utrecht, and acknowledge the financial support of the
Spinoza grant from NWO. The second author (FZ) thanks S. Cosentino for
their clarifying comments in relation to symbolic dynamics. This work
is supported in part by the DGAPA-UNAM grant IN105197.

\vfill\eject

\section{References}

\noindent \item{[1]} Coolen, A.C.C. and Sherrington, D. {\it
Competition between Pattern Reconstruction and Sequence Processing in
Non-symmetric Neural Networks}. J. Phys. A: Math. Gen. {\bf 25} (1992)
5493-5526.

\noindent \item{[2]} Crisanti A., Falcioni M. and Vulpiani A. {\it
Transition from Regular to Complex Behavior in a Discrete
Deterministic Asymmetric Neural Network Model}. J. Phys. A:
Math. Gen. {\bf 26} (1993) 3441.

\noindent \item{[3]} Coolen A.C.C. and Ruijgrok Th. W. {\it Image
Evolution in Hopfield Networks}. Phys. Rev. A {\bf 38} (1988)
4253-4255.

\noindent \item{[4]} Zertuche F., L\'opez-Pe\~na R. and Waelbroeck H.
{\it Recognition of Temporal Sequences of Patterns with
State-Dependent Synapses}. J. Phys. A: Math. Gen. {\bf 27} (1994)
5879-5887.

\noindent \item{[5]} Kauffman, S.  {\it The Origins of Order:
Self-Organization and Selection in Evolution}. (Oxford University
Press, New York) (1993).

\noindent \item{[6]} Derrida, B. {\it Dynamical Phase Transitions in
Random Networks of Automata}, in: {\it Chance and Matter} edited by
J. Souletie, J. Vannimenus and R. Stora (North Holland) (1987).

\noindent \item{[7]} Weisbuch, G. {\it Complex Systems Dynamics}.
(Addison Wesley, Redwood City, CA) (1991).

\noindent \item{[8]} Chaitin, G.J. {\it Randomness and Complexity in
Pure Mathematics}. Int. J. Bifurcation and Chaos {\bf 4} (1994) 3-15;
Chaitin, G.J. in: {\it Guanajuato Lectures on Complex Systems and
Binary Networks}. Springer Verlag Lecture Notes series. Eds. R.
L\'opez Pe\~na, R. Capovilla, R. Garc\'\i a-Pelayo, H. Waelbroeck and
F. Zertuche. (1995).

\noindent \item{[9]} Hao, B.-L. {\it Elementary Symbolic Dynamics and
Chaos in Dissipative Systems} (World Scientific, Singapore) (1989);
Bruin, H.  {\it Combinatorics of the Kneading Map}.  Int. J.
Bifurcation and Chaos {\bf 5} (1995) 1339-1349.

\noindent \item{[10]} Devaney R. {\it An Introduction to Chaotic
Dynamical Systems}, Addison Wesley Publ. Co. Reading MA, (1989);
Katok A. and Hasselblatt B. {\it Introduction to the Modern Theory of
Dynamical Systems}. Cambridge University Press. Cambridge (1995);
Robinson C. {\it Dynamical Systems}. CRC Press. Boca Raton FL (1995).

\noindent \item{[11]} Wiggins, S.  {\it Dynamical Systems and Chaos}.
(Springer-Verlag, New York) (1990). 

\noindent \item{[12]} Wiggins, S.  {\it Global Bifurcations and
Chaos}.  (Springer-Verlag, New York) (1988). 

\noindent \item{[13]} D. Lind and B. Marcus {\it An Introduction to
Symbolic Dynamics and Coding}. (Cambridge University Press) (1995).

\noindent \item{[14]} Munkres, J.R.  {\it Topology a First Course}.
(Prentice-Hall, New Jersey) (1975).

\noindent \item{[15]} Buescu J. and Stewart I. {\it Liapunov Stability
and Adding Machines.}  Ergod. Th. \& Dynam. Sys. {\bf 15} (1995)
271-290; Buescu J. {\it Exotic Attractors, Liapunov Stability and
Riddled  Basins.} Progress in Mathematics {\bf 153} (Birkh\"auser
Verlag, Basel) (1997). 

\noindent \item{[16]} Moser J. {\it Stable and Random Motions in
Dynamical Systems}. (Princeton University Press, Princeton) (1973);
similar criteria were given in: Alekseev, V. M. {\it Quasirandom
dynamical systems, I-III.} Math. USSR-Sb. {\bf 5} (1968) 73-128; {\bf
6} (1968) 505-560; {\bf 7} (1969) 1-43.

\noindent \item{[17]} Sompolinsky H. and Kanter I. {\it Temporal
Association in Asymmetric Neural Networks}. Phys. Rev. Lett. {\bf 57}
(1986) 2861; Riedel U., K\"uhn R. and van Hemmen J.L. {\it Temporal
Sequences and Chaos in Neural Nets}. Phys. Rev. A {\bf 38} (1988)
1105. 

\noindent \item{[18]} Grassberger P. and Procaccia I. Phys. D {\bf
9} (1983) 189-208; Phys. Rev. Lett. {\bf 50} (1983) 346-349.

\noindent \item{[19]} Lorenz, E.N. {\it Dimension of Weather and
Climate Attractors}. Nature {\bf 353} (1991) 241, and references
therein. 

\noindent \item{[20]} Ruijgrok, Th. W., personal communication.

\vfill \eject

\section{Figure Captions}

\noindent \item{[1]} Construction of the Cantor sets $ \Xi_i $, $ i =
1, \dots, N $ by suppressing from $ \left[ 0, 1 \right] $ the open
middle $ 1 / a_i $ part, $ 1 < a_i < \infty $. The remaining $ 2^n $
intervals at the $ n^{th} $ step of the construction are of length $
l_n = {1 \over 2^n} \left( 1 - {1 \over a_i} \right)^n $.

\noindent \item{[2]} The number of times a $k=4$ random cellular
automata with $N=200$ returns to a Hamming distance $d_H$ of a point
half-way along the trajectory is represented as a function of $d_H$
(solid dots). The best-fitting Gaussian is also given for comparison
(open dots).

\noindent \item{[3]} The number of returns to Hamming distance $d_H$
is shown, for an asymmetric neural network with $N=100$ (solid dots).
The best-fitting Gaussian is also given for comparison (open dots). 

\noindent \item{[4]} The number of returns to Hamming distance $d_H$
is given for a neural network model with memory, with $N = 49$.
Unlike the previous examples, which correspond to dynamical systems
without memory, we find many analogues. There is one return with $d_H
= 0$; the system did not fall on a limit-cycle at that point because
the dynamics also considers binary words further back in time.

\noindent \item{[5]} The correlation graph $N(\rho)$ gives the
effective attractor dimension for the neural network with memory,
$D_a \approx 2$ in the range $0.002 < \rho < 0.02$. The distance
$\rho$ is given by equation (3).

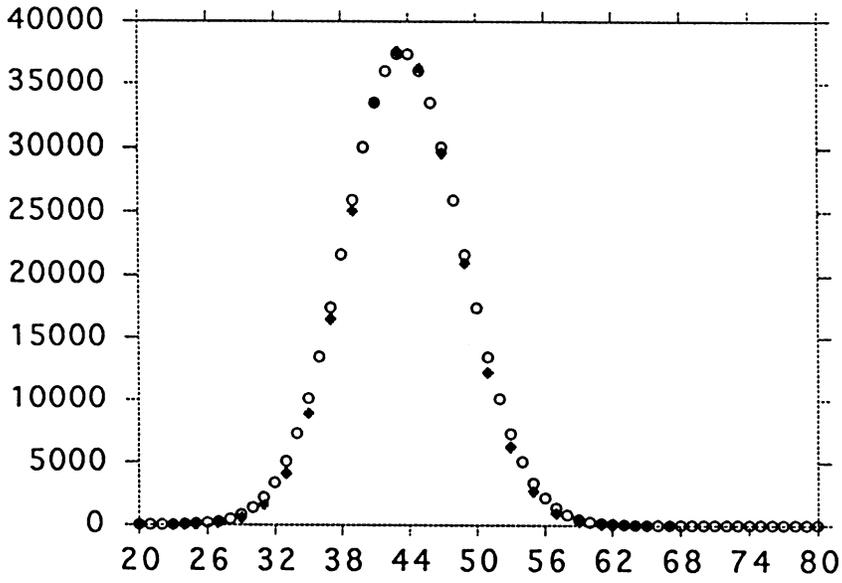
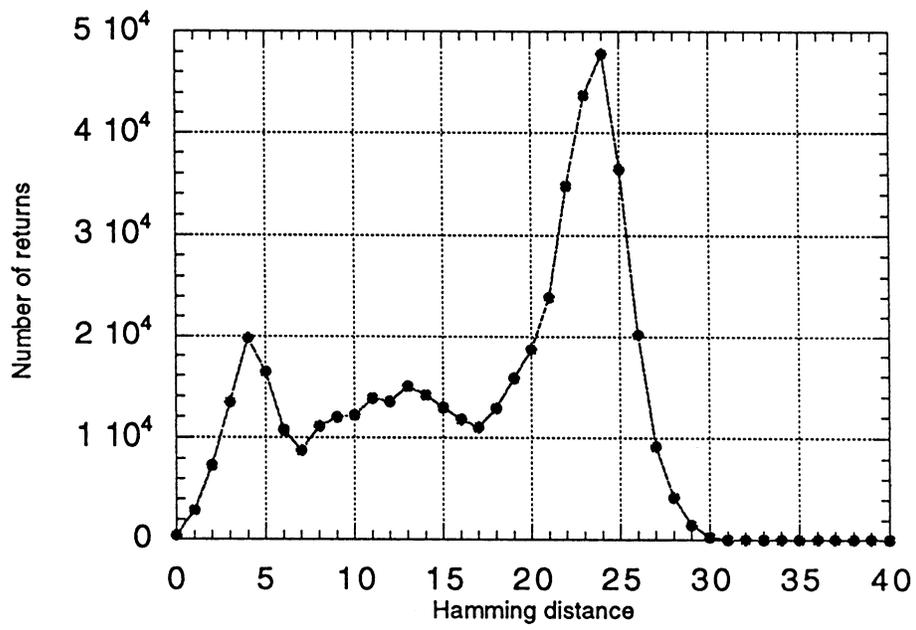
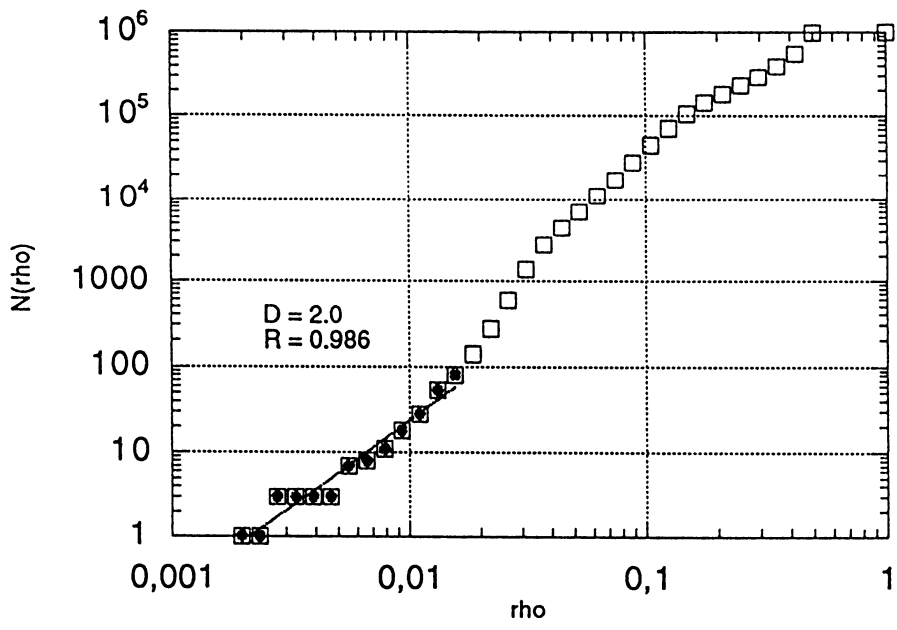
\end